\documentclass[10pt]{IEEEtran15}
\usepackage{graphicx}
\usepackage{amsmath}
\usepackage{amssymb}
\usepackage{color}
\usepackage{epsfig}

\newcommand{\Define}{\stackrel{\triangle}{=}}

\begin{document}
\twocolumn

\title{\LARGE Low-Complexity Near-ML Decoding of Large Non-Orthogonal 
STBCs Using PDA}
\author{Saif K. Mohammed, A. Chockalingam, and B. Sundar Rajan \\
\vspace{-0mm}
{\normalsize Department of ECE, Indian Institute of Science,
Bangalore 560012, INDIA} 
\vspace{-10mm}
}
\maketitle
\thispagestyle{empty}

\begin{abstract}
Non-orthogonal space-time block codes (STBC) from cyclic division 
algebras (CDA) having large dimensions are attractive because they can 
simultaneously achieve 
both high spectral efficiencies (same spectral efficiency as in V-BLAST 
for a given number of transmit antennas) {\em as well as} full transmit 
diversity. Decoding of non-orthogonal STBCs with hundreds of dimensions 
has been a challenge. In this paper, we present a probabilistic data 
association (PDA) based algorithm for decoding non-orthogonal STBCs 
with large dimensions. Our simulation results show that the proposed 
PDA-based algorithm achieves near SISO AWGN uncoded BER as well as 
near-capacity coded BER (within about 5 dB of the theoretical capacity) 
for large non-orthogonal STBCs from CDA. We study the effect of spatial 
correlation on the BER, and show that the performance loss due to spatial 
correlation can be alleviated by providing more receive spatial dimensions. 
We report good BER performance when a training-based iterative 
decoding/channel estimation is used (instead of assuming perfect 
channel knowledge) in channels with large coherence times. A comparison 
of the performances of the PDA algorithm and the likelihood ascent search 
(LAS) algorithm (reported in our recent work) is also presented.
\end{abstract} 

\vspace{-4.0mm}

\vspace{-4.0mm}
\section{Introduction}
\label{sec1}
\vspace{-1.0mm}
Multiple-input multiple-output (MIMO) systems that employ non-orthogonal 
space-time block codes (STBC) from cyclic division algebras (CDA) for 
arbitrary number of transmit antennas, $N_t$, are quite attractive because 
they can simultaneously provide both {\em full-rate} (i.e., $N_t$ complex 
symbols per channel use, which is same as in V-BLAST) as well as {\em full 
transmit diversity} \cite{bsr}. The $2\times 2$ Golden code is a well 
known non-orthogonal STBC from CDA for 2 transmit antennas \cite{gold05}. 
High spectral efficiencies of the order of tens of bps/Hz can be achieved 
using large non-orthogonal STBCs. For example, a $16\times 16$ STBC from 
CDA has 256 complex symbols in it with 512 real dimensions; with 16-QAM 
and rate-3/4 turbo code, this system offers a high spectral efficiency 
of 48 bps/Hz. Decoding of non-orthogonal STBCs with such large dimensions, 
however, has been a challenge. Sphere decoder and its low-complexity 
variants are prohibitively complex for decoding such STBCs with hundreds 
of dimensions. 

In this paper, we present a probabilistic data association (PDA) based 
algorithm for decoding large non-orthogonal STBCs from CDA. Key attractive 
features of this algorithm are its low-complexity and near-ML performance 
in systems with large dimensions (e.g., hundreds of dimensions). While 
creating hundreds of dimensions in space alone (e.g., V-BLAST) requires
hundreds of antennas, use of non-orthogonal STBCs from CDA can create 
hundreds of dimensions with just tens of antennas (space) and tens of 
channel uses (time). Given that 802.11 smart WiFi products with 12 
transmit antennas\footnote{12 antennas in these products are now used 
only for beamforming. Single-beam multi-antenna approaches can offer
range increase and interference avoidance, but not spectral efficiency 
increase.} at 2.5 GHz are now commercially 
available \cite{ruckus} (which establishes that issues related to 
placement of many antennas and RF/IF chains can be solved in large 
aperture communication terminals like set-top boxes/laptops), large 
\hspace{-1.5mm} non-orthogonal STBCs (e.g., $16\times 16$ STBC from 
CDA) in combination with large dimension near-ML decoding using PDA can 
enable communications at increased spectral efficiencies of the order of 
tens of bps/Hz (note that current standards achieve only $< 10$ bps/Hz 
using only up to 4 transmit antennas).

PDA, originally developed for target tracking, is widely used in digital 
communications \cite{pda1}-\cite{pda9}. Particularly, PDA algorithm is a 
reduced complexity alternative to the a posteriori probability (APP) 
decoder/detector/equalizer. Near-optimal performance has been demonstrated 
for PDA-based multiuser detection in CDMA systems \cite{pda1}-\cite{pda3}. 
PDA has been used in the detection of V-BLAST signals with small 
number of dimensions \cite{pda5}-\cite{pda9}. To our knowledge, PDA has not 
been reported for {\em decoding non-orthogonal STBCs with hundreds of 
dimensions} so far. Our results in this paper can be summarized as follows:
\begin{itemize}
\item 	We adapt the PDA algorithm for decoding non-orthogo\-nal STBCs 
	with large dimensions. With i.i.d fading and perfect CSIR,
        the algorithm achieves near-SISO AWGN uncoded BER and 
	near-capacity coded BER (within about 5 dB of the theoretical 
	capacity) for $12\times 12$ STBC from CDA, 4-QAM, 
	rate-3/4 turbo code, and 18 bps/Hz.
\item	Relaxing the perfect CSIR assumption, we report
	results with a training based iterative PDA decoding/cha\-nnel
	estimation scheme. The iterative scheme is shown to be
	effective with large coherence times.
\item   Relaxing the i.i.d fading assumption by adopting a 
	spatially correlated MIMO channel model (proposed by Gesbert 
	et al in \cite{gesbert}), we show that the performance 
	loss due to spatial correlation is alleviated by using more 
	receive spatial dimensions for a fixed receiver aperture. 
\item	Finally, the performance of the PDA algorithm is compared with 
	that of the likelihood ascent search (LAS) algorithm we recently 
	presented in \cite{jsac}-\cite{gcom08}. The PDA algorithm is 
	shown to perform 
	better than the LAS algorithm at low SNRs for higher-order QAM 
	(e.g., 16-QAM), and in the presence of spatial correlation.
\end{itemize}

\vspace{-3.00mm}
\section{System Model}
\label{sec2}
\vspace{-1.0mm}
Consider a STBC MIMO system with multiple transmit
and receive antennas. An $(n,p,k)$ STBC is
represented by a matrix {\small ${\bf X}_c \in {\mathbb C}^{n \times p}$},
where $n$ and $p$ denote the number of transmit antennas and number of 
time slots, respectively, and $k$ denotes the number of complex data 
symbols sent in one STBC matrix. The $(i,j)$th entry in ${\bf X}_c$ 
represents the complex number transmitted from the $i$th transmit 
antenna in the $j$th time slot. The rate of an STBC is $\frac{k}{p}$.
Let $N_r$ and $N_t=n$ denote the number of receive and transmit antennas,
respectively. Let ${\bf H}_c \in {\mathbb C}^{N_r\times N_t}$ denote the 
channel gain matrix, where the $(i,j)$th entry in ${\bf H}_c$ is the 
complex channel gain from the $j$th transmit antenna to the $i$th receive
antenna. We assume that the channel gains remain constant over one STBC 
matrix and vary (i.i.d) from one STBC matrix  to the other. Assuming rich 
scattering, we model the entries of ${\bf H}_c$ as i.i.d 
$\mathcal C \mathcal N(0,1)$. The received space-time signal matrix, 
{\small ${\bf Y}_c \in {\mathbb C}^{N_r \times p}$}, can be written as
\begin{equation}
\label{SystemModel}
{\bf Y}_c = {\bf H}_c{\bf X}_c + {\bf N}_c,
\end{equation}
where ${\bf N}_c \in {\mathbb C}^{N_r \times p}$ is the noise matrix 
at the receiver and its entries are modeled as i.i.d
$\mathcal C \mathcal N\big(0,\sigma^2=\frac{N_tE_s}{\gamma}\big)$,
where $E_s$ is the average energy of the transmitted symbols, and
$\gamma$ is the average received SNR per receive antenna \cite{jafarkhani},
and the $(i,j)$th entry in ${\bf Y}_c$ is the received signal at the $i$th
receive antenna in the $j$th time-slot. Consider linear dispersion STBCs, 
where ${\bf X}_c$ can be written in the form \cite{jafarkhani}
\begin{eqnarray}
\label{SystemModelStbcLpx}
{\bf X}_c & = & \sum_{i = 1}^{k} x_c^{(i)} {\bf A}_c^{(i)},
\end{eqnarray}
where $x_c^{(i)}$ is the $i$th complex data symbol, and
${\bf A}_c^{(i)} \in {\mathbb C}^{N_t \times p}$ is its 
corresponding weight matrix. The received signal model in 
(\ref{SystemModel}) can be written in an equivalent V-BLAST 
form as
\begin{equation}
\label{SystemModelvec2}
{\bf y}_c \,\, = \,\, \sum_{i=1}^{k} x_c^{(i)}\, (\widehat{{\bf H}}_c\, {\bf a}_c^{(i)}) + {\bf n}_c \,\, = \,\, \widetilde{{\bf H}}_c {\bf x}_c + {\bf n}_c,
\end{equation}
where
${\bf y}_c \in {\mathbb C}^{N_rp \times 1} = vec\,({\bf Y}_c)$,
$\widehat{{\bf H}}_c \in {\mathbb C}^{N_rp \times N_tp} = ({\bf I} \otimes {\bf H}_c)$,
${\bf a}_c^{(i)} \in {\mathbb C}^{N_tp \times 1} = vec\,({\bf A}_c^{(i)})$,
${\bf n}_c \in {\mathbb C}^{N_rp \times 1} = vec\,({\bf N}_c)$,
${\bf x}_c \in {\mathbb C}^{k \times 1}$
whose $i$th entry is the data symbol $x_c^{(i)}$, and
$\widetilde{{\bf H}}_c \in {\mathbb C}^{N_rp \times k}$ whose
$i$th column is
$\widehat{{\bf H}}_c \, {\bf a}_c^{(i)}$, $i=1,2,\cdots,k$.
Each element of ${\bf x}_c$ is an $M$-PAM/$M$-QAM symbol. 
Let ${\bf y}_c$, $\widetilde{{\bf H}}_c$, ${\bf x}_c$, ${\bf n}_c$
be decomposed into real and imaginary parts as: 
\begin{eqnarray}
\label{SystemModelDecompose}
{\bf y}_c={\bf y}_I + j{\bf y}_Q, &  {\bf x}_c = {\bf x}_I + j{\bf x}_Q,
\nonumber \\
{\bf n}_c={\bf n}_I+j{\bf n}_Q, & \widetilde{{\bf H}}_c={\bf H}_I+j{\bf H}_Q.
\end{eqnarray}
Further, we define
{\normalsize
${\bf H}_r \in {\mathbb R}^{2N_rp \times 2k}$,
${\bf y}_r \in {\mathbb R}^{2N_rp \times 1}$,
${\bf x}_r \in {\mathbb R}^{2k \times 1}$}, and
{\normalsize
${\bf n}_r \in {\mathbb R}^{2N_rp \times 1}$} as
\begin{eqnarray}
\label{SystemModelRealDef}
{\bf H}_r = \left(\begin{array}{cc}{\bf H}_I \hspace{2mm} -{\bf H}_Q \\
{\bf H}_Q  \hspace{5mm} {\bf H}_I \end{array}\right),
\hspace{4mm}
{\bf y}_r = [{\bf y}_I^T \hspace{2mm} {\bf y}_Q^T ]^T, \\
\hspace{4mm}
{\bf x}_r = [{\bf x}_I^T \hspace{2mm} {\bf x}_Q^T ]^T,
\hspace{4mm}
{\bf n}_r = [{\bf n}_I^T \hspace{2mm} {\bf n}_Q^T ]^T.
\end{eqnarray}

\thanks{
{\small
\line(1,0){505}
\[
\label{eqn}
\hspace{1.3cm}
\left[
\begin{array}{ccccc}
\sum_{i=0}^{n-1}d_{0,i}\,t^i & \delta\sum_{i=0}^{n-1}d_{n-1,i}\,\omega_n^i\,t^i & \delta\sum_{i=0}^{n-1}d_{n-2,i}\,\omega_n^{2i}\,t^i & \cdots & \delta\sum_{i=0}^{n-1}d_{1,i}\,\omega_n^{(n-1)i}\,t^i \\
\sum_{i=0}^{n-1}d_{1,i}\,t^i & \sum_{i=0}^{n-1}d_{0,i}\,\omega_n^i\,t^i & \delta\sum_{i=0}^{n-1}d_{n-1,i}\,\omega_n^{2i}\,t^i & \cdots & \delta\sum_{i=0}^{n-1}d_{2,i}\,\omega_n^{(n-1)i}\,t^i \\
\sum_{i=0}^{n-1}d_{2,i}\,t^i & \sum_{i=0}^{n-1}d_{1,i}\,\omega_n^i\,t^i & \sum_{i=0}^{n-1}d_{0,i}\,\omega_n^{2i}\,t^i & \cdots & \delta\sum_{i=0}^{n-1}d_{3,i}\,\omega_n^{(n-1)i}\,t^i \\
\vdots & \vdots & \vdots & \vdots & \vdots \\
\sum_{i=0}^{n-1}d_{n-2,i}\,t^i & \sum_{i=0}^{n-1}d_{n-3,i}\,\omega_n^i\,t^i & \sum_{i=0}^{n-1}d_{n-4,i}\,\omega_n^{2i}\,t^i & \cdots & \delta \sum_{i=0}^{n-1}d_{n-1,i}\,\omega_n^{(n-1)i}t^i \\
\sum_{i=0}^{n-1}d_{n-1,i}\,t^i & \sum_{i=0}^{n-1}d_{n-2,i}\,\omega_n^i\,t^i & \sum_{i=0}^{n-1}d_{n-3,i}\,\omega_n^{2i}\,t^i & \cdots & \sum_{i=0}^{n-1}d_{0,i}\,\omega_n^{(n-1)i}\,t^i
\end{array}
\right]. \hspace{10mm} (\mbox{9.a})
\]
}
}

\vspace{-5mm}
Now, (\ref{SystemModelvec2}) can be written as
\begin{eqnarray}
\label{SystemModelReal}
{\bf y}_r & = & {\bf H}_r{\bf x}_r + {\bf n}_r.
\end{eqnarray}
Henceforth, we work with the real-valued system in
(\ref{SystemModelReal}). For notational simplicity, we drop
subscripts $r$ in (\ref{SystemModelReal}) and write
\begin{eqnarray}\label{SystemModelII}
{\bf y} & = & {\bf H}' {\bf x} + {\bf n},
\end{eqnarray}
where {\small ${\bf H}' = {\bf H}_r \in {\mathbb R}^{2N_rp \times 2k}$,
${\bf y} = {\bf y}_r \in {\mathbb R}^{2N_rp \times 1}$,
${\bf x} = {\bf x}_r \in {\mathbb R}^{2k \times 1}$}, and
{\small ${\bf n} = {\bf n}_r \in {\mathbb R}^{2N_rp \times 1}$.}
We assume that the channel coefficients are known at the receiver 
but not at the transmitter. Let $\mathbb A_i$ denote the $M$-PAM 
signal set from which $x_i$ ($i$th entry of ${\bf x}$) takes values, 
$i=0,\cdots,2k-1$. Now, define a $2k$-dimensional signal space 
$\mathbb S$ to be the Cartesian product of $\mathbb A_0$ to 
$\mathbb A_{2k-1}$. The ML solution is then given by
\begin{eqnarray}
\label{MLdetection}
{\bf d}_{ML} & = & {\mbox{arg min}\atop{{\bf d} \in {\mathbb S}}}
{\bf d}^T ({\bf H}')^T{\bf H}'{\bf d} - 2{\bf y}^T{\bf H}'{\bf d},
\end{eqnarray}
whose complexity is exponential in $k$.

\vspace{-3.5mm}
\subsection{Full-rate Non-orthogonal STBCs from CDA}
\vspace{-1.0mm}
We focus on the detection of square (i.e., $
n\hspace{-0.9mm}=\hspace{-0.9mm}p\hspace{-0.9mm}=\hspace{-0.9mm}N_t$),
full-rate (i.e., 
$k\hspace{-0.9mm}=\hspace{-0.9mm}pn\hspace{-0.9mm}=\hspace{-0.9mm}N_t^2$),
circulant (where the weight matrices ${\bf A}_c^{(i)}$'s are permutation 
type), non-orthogonal STBCs from CDA \cite{bsr}, whose construction for 
arbitrary number of transmit antennas $n$ is given by the matrix in 
Eqn.(9.a) given at the bottom of this page. In (9.a), 
{\small $\omega_n=e^{\frac{{\bf j}2\pi}{n}}$, ${\bf j}=\sqrt{-1}$, 
and $d_{u,v}$, $0\leq u,v \leq n-1$} are the $n^2$ data symbols from a QAM 
alphabet. When $\delta=t=1$, the code in (9.a) is information lossless 
(ILL), and when $\delta=e^{\sqrt{5}\,{\bf j}}$ and $t=e^{{\bf j}}$, it is 
of full-diversity and information lossless (FD-ILL) \cite{bsr}.
High spectral efficiencies with large $n$ can be achieved using this 
code construction. However, since these STBCs are non-orthogonal, 
ML detection gets increasingly impractical for large $n$. Consequently, 
a key challenge in realizing the benefits of these large STBCs in 
practice is that of achieving near-ML performance for large $n$ at 
low decoding complexities. The BER performance results we report in 
Sec. \ref{sec4} show that the PDA-based decoding algorithm we propose 
in the following section essentially meets this challenge.

\vspace{-3mm}
\section{Proposed PDA-Based Decoding}
\label{sec3}
\vspace{-1.0mm}
In this section, we present the proposed PDA-based decoding algorithm 
for square QAM. The applicability of the algorithm to any rectangular 
QAM is straightforward. In the real-valued system model in 
(\ref{SystemModelII}), each entry of ${\bf x}$ belongs to a 
$\sqrt{M}$-PAM constellation, where $M$ is the size of the original 
square QAM constellation. Let $b_i^{(0)},b_i^{(1)},\cdots,b_i^{(q-1)}$ 
denote the $q=\log_2(\sqrt{M})$ constituent bits of the $i$th entry 
$x_i$ of ${\bf x}$. 

\newpage
We can write the value of each entry of ${\bf x}$ 
as a linear combination of its constituent bits as
\begin{eqnarray}
\label{linearComb}
x_i&=&\sum_{j=0}^{q-1} 2^j \, b_i^{(j)}, \,\,\,\,\,\,\,\, i=0,1,\cdots,2k-1.  
\end{eqnarray}
Let ${\bf b} \in \{+1, -1\}^{2qk \times 1}$, defined as 
\begin{equation}
\hspace{-0.0mm}
{\bf b} \,\, \Define \,\, \left[ b_0^{(0)} \cdots b_0^{(q-1)} b_1^{(0)} \cdots b_1^{(q-1)} \cdots b_{2k-1}^{(0)} \cdots b_{2k-1}^{(q-1)} \right]^T\hspace{-1mm},  \hspace{-0mm}
\end{equation}
denote the transmitted bit vector. Defining 
{\small ${\bf c} \Define [2^{0} \, 2^{1} \cdots 2^{q-1}]$},
we can write ${\bf x}$ as
\begin{eqnarray}
\label{b2seq}
{\bf x} & = & ({\bf I} \otimes {\bf c}) {\bf b}, 
\end{eqnarray}
where ${\bf I}$ is the $2k \times 2k$ identity matrix.
Using (\ref{b2seq}), we can rewrite (\ref{SystemModelII}) as
\begin{eqnarray}
\label{revSystemModelII}
{\bf y} & = & \underbrace{{\bf H}'({\bf I} \otimes {\bf c})}_{{\Define \,\,\bf H}} \, {\bf b} +  {\bf n}, 
\end{eqnarray}
where ${\bf H} \in {\mathbb R}^{2N_rp  \times 2qk}$ is the effective 
channel matrix. Our goal is to obtain $\widehat{{\bf b}}$, an 
estimate of the ${\bf b}$ vector. For this, we iteratively update 
the statistics of each bit of ${\bf b}$, as described in 
the following subsection, for a certain number of iterations, and hard 
decisions are made on the final statistics to get $\widehat{{\bf b}}$.

\vspace{-3.5mm}
\subsection{Iterative Procedure} 
\label{sec3p1}
\vspace{-1mm}
The algorithm is iterative in nature, where $2qk$ statistic 
updates, one for each of the constituent bits, are performed 
in each iteration. We start the algorithm by initializing the 
a priori probabilities as {\small $P(b_i^{(j)}=+1) = P(b_i^{(j)}=-1) = 0.5$},
$\forall \, i=0,\cdots,2k-1$ and $j=0,\cdots,q-1$. In an iteration,
the statistics of the bits are updated sequentially, i.e., the ordered 
sequence of updates in an iteration is 
$\big\{b_0^{(0)},\cdots,b_0^{(q-1)},\cdots\cdots,\\  
b_{2k-1}^{(0)},\cdots b_{2k-1}^{(q-1)} \big\}$. 
The steps involved in each iteration of the algorithm are derived
as follows.

The likelihood ratio of bit $b_i^{(j)}$ in an iteration, denoted by 
$\Lambda_{i}^{(j)}$, is given by
\begin{eqnarray}
\label{llr_i_j_k} 
\Lambda_{i}^{(j)} & \Define & \frac{P\big(b_i^{(j)} = +1 | {\bf y}\big)}{P\big(b_i^{(j)} = -1 | {\bf y}\big)}  \nonumber \\
& = & 
\underbrace{\frac{P\big({\bf y} | b_i^{(j)} = +1\big)}{P\big({\bf y} | b_i^{(j)} = -1\big)}}_{\Define \,\, \beta_{i}^{(j)}} \,\, \underbrace{\frac{P\big(b_i^{(j)} = +1\big)}{P\big(b_i^{(j)} = -1\big)}}_{\Define \,\, \alpha_{i}^{(j)}}. 
\end{eqnarray}
Denoting the $t$th column of ${\bf H}$ by ${\bf h}_t$, we can write
(\ref{revSystemModelII}) as 
\begin{eqnarray}
\label{ext_1} 
\hspace{-8mm}
{\bf y} & = & {\bf h}_{qi+j} \, b_i^{(j)} + \underbrace{\sum_{l=0}^{2k-1} \sum_{{\stackrel{m=0}{m\neq q(i-l)+j}}}^{q-1} \hspace{-3mm} {\bf h}_{ql+m} \, b_l^{(m)} + {\bf n}}_{\Define \,\, \widetilde{{\bf n}}}, 
\end{eqnarray}
where $\widetilde{{\bf n}} \in {\mathbb R}^{2N_rp \times 1}$ is the 
interference plus noise vector. To calculate 
$\beta_{i}^{(j)}$, we approximate the distribution of $\widetilde{{\bf n}}$ 
to be Gaussian, and hence ${\bf y}$ is Gaussian conditioned on $b_i^{(j)}$.
Since there are $2qk-1$ terms in the double summation in (\ref{ext_1}), 
this Gaussian approximation gets increasingly accurate for large $N_t$
(note that $k=N_t^2$). Since a Gaussian distribution is fully characterized 
by its mean and covariance, we evaluate the mean and covariance of ${\bf y}$ 
given $b_i^{(j)} = +1$ and $b_i^{(j)} = -1$. For notational simplicity, let 
us define $p_{i}^{j+} \Define P(b_i^{(j)}=+1)$ and 
$p_{i}^{j-} \Define P(b_i^{(j)}=-1)$. It is clear that 
$p_{i}^{j+} + p_{i}^{j-} = 1$.

Let 
${\mbox{\boldmath{$\mu$}}}_i^{j+} \Define {\mathbb E}({\bf y}|b_i^{(j)}=+1)$
and
${\mbox{\boldmath{$\mu$}}}_i^{j-} \Define {\mathbb E}({\bf y}|b_i^{(j)}=-1)$, 
where ${\mathbb E}(.)$ denotes the expectation operator. Now, from 
(\ref{ext_1}), we can write ${\mbox{\boldmath{$\mu$}}}_i^{j+}$ as
\begin{eqnarray}
\label{Exp1}
\hspace{-7mm} 
{\mbox{\boldmath{$\mu$}}}_i^{j+} 
& = & {\bf h}_{qi+j} + \sum_{l=0}^{2k-1} \hspace{-1mm} \sum_{{\stackrel{m=0}{m\neq q(i-l)+j}}}^{q-1}\hspace{-5mm} {\bf h}_{ql+m} (2p_{l}^{m+} - 1). 
\end{eqnarray}
Similarly, we can write ${\mbox{\boldmath{$\mu$}}}_i^{j-}$ as
\begin{eqnarray}
\label{Exm1} 
\hspace{-0mm}
{\mbox{\boldmath{$\mu$}}}_i^{j-} 
& = & -{\bf h}_{qi+j} + \hspace{-1mm} \sum_{l=0}^{2k-1} \hspace{-1mm} \sum_{{\stackrel{m=0}{m\neq q(i-l)+j}}}^{q-1}\hspace{-5mm} {\bf h}_{ql+m} (2p_{l}^{m+} - 1)  \nonumber \\ 
& = & {\mbox{\boldmath{$\mu$}}}_i^{j+} - 2{\bf h}_{qi+j}. 
\end{eqnarray}

\vspace{-3mm}
Next, the {\small $2N_rp\times 2N_rp$} covariance matrix 
${\bf C}_{i}^{j}$ of ${\bf y}$ given $b_i^{(j)}$ is given by 
\begin{eqnarray}
\label{covpm1} 
\hspace{-4mm}
{\bf C}_{i}^{j} & = &
{\mathbb E}\bigg\{ \Big[ {\bf n} + \hspace{-1mm} 
\sum_{l=0}^{2k-1} \hspace{-1mm} \sum_{{\stackrel{m=0}{m\neq q(i-l)+j}}}^{q-1}\hspace{-4mm} {\bf h}_{ql+m} (b_l^{(m)} -  2p_l^{m+} + 1) \Big] \nonumber \\ 
& & \hspace{-7mm} \Big[ {\bf n} + \hspace{-1mm} 
\sum_{l=0}^{2k-1} \hspace{-1mm} \sum_{{\stackrel{m=0}{m\neq q(i-l)+j}}}^{q-1}\hspace{-3mm}
{\bf h}_{ql+m} (b_l^{(m)} - 2p_l^{m+}+ 1)\Big]^T\bigg\}. 
\end{eqnarray}
Assuming independence among the constituent bits, we can simplify 
${\bf C}_{i}^{j}$ in (\ref{covpm1}) as 
\begin{eqnarray}
\label{covpm2}
{\bf C}_{i}^{j} \, = \, \sigma^2 {\bf I} \, + \hspace{-0mm} 
\sum_{l=0}^{2k-1} \hspace{-1mm} \sum_{{\stackrel{m=0}{m\neq q(i-l)+j}}}^{q-1}\hspace{-4mm}
{\bf h}_{ql+m}\, {\bf h}_{ql+m}^T\, 4 p_{l}^{m+} (1 - p_{l}^{m+}).
\end{eqnarray}
Using the above mean and covariance expressions, we can write the 
distribution of ${\bf y}$ given $b_i^{(j)} = \pm 1$ as
\begin{eqnarray}
\hspace{-5mm}
\label{pdf_p1}
P({\bf y} | b_i^{(j)} = \pm 1) & = & 
\frac{e^{- ({\bf y} - {\mbox{\boldmath{$\mu$}}}_i^{j\pm})^T ({\bf C}_{i}^{j})^{-1} ({\bf y} - {\mbox{\boldmath{$\mu$}}}_{i}^{j\pm})}} {(2\pi)^{N_rp} \vert {\bf C}_{i}^{j} \vert^{\frac{1}{2}} }.  
\end{eqnarray}
Similarly, $P({\bf y} | b_i^{(j)} = -1)$ is given by 
\begin{eqnarray}
\label{pdf_m1}
\hspace{-3mm}
P({\bf y} | b_i^{(j)} = -1) & = & 
\frac{e^{-({\bf y} - {\mbox{\boldmath{$\mu$}}}_{i}^{j-})^T ({\bf C}_{i}^{j})^{-1} ({\bf y} - {\mbox{\boldmath{$\mu$}}}_{i}^{j-})}} {(2\pi)^{N_rp} \vert {\bf C}_{i}^{j} \vert^{\frac{1}{2}} }.  
\end{eqnarray}
Using (\ref{pdf_p1}) and (\ref{pdf_m1}), $\beta_i^{j}$ can be written as

\vspace{-5mm}
{\small
\begin{eqnarray}
\label{ext_fin1} 
\hspace{-4mm}
\beta_{i}^{j} & = & \frac{P({\bf y} | b_i^{(j)} = +1)}{P({\bf y} | b_i^{(j)} = -1)} \nonumber \\ 
& \hspace{-19mm} = & \hspace{-12mm} 
 e^{-\left( ({\bf y} - {\mbox{\boldmath{$\mu$}}}_{i}^{j+})^T ({\bf C}_{i}^{{j}})^{-1} ({\bf y} - {\mbox{\boldmath{$\mu$}}}_{i}^{j+}) - ({\bf y} - {\mbox{\boldmath{$\mu$}}}_{i}^{j-})^T ({\bf C}_{i}^{j})^{-1} ({\bf y} - {\mbox{\boldmath{$\mu$}}}_{i}^{j-})\right)}\hspace{-0.5mm}. \hspace{-0.5mm}
\end{eqnarray}
}

\vspace{-6mm}
Using $\alpha_i^{(j)}$ and $\beta_i^{(j)}$, 
$\Lambda_i^{(j)}$ is computed using (\ref{llr_i_j_k}). 
Now, using the value of $\Lambda_i^{(j)}$, the statistics
of $b_i^{(j)}$ is updated as follows. From (\ref{llr_i_j_k}), and 
using {\small $P(b_i^{(j)}=+1|{\bf y}) + P(b_i^{(j)}=-1|{\bf y}) = 1$},
we have 
\begin{eqnarray}
P(b_i^{(j)} = +1|{\bf y}) \, = \, \frac{\Lambda_i^{(j)}}{1+\Lambda_i^{(j)}},
\end{eqnarray}

\vspace{-5mm}
and

\vspace{-5mm}
\begin{eqnarray}
P(b_i^{(j)}=-1|{\bf y}) & = & \frac{1}{1+\Lambda_i^{(j)}}.
\end{eqnarray}
As an approximation, dropping the conditioning on ${\bf y}$, 
\begin{eqnarray}
\label{appx1}
P(b_i^{(j)}=+1) \, \approx \, \frac{\Lambda_i^{(j)}}{1+\Lambda_i^{(j)}},
\end{eqnarray}

\vspace{-4mm}
and

\vspace{-6mm}
\begin{eqnarray}
\label{appx2}
P(b_i^{(j)}=-1) & \approx & \frac{1}{1+\Lambda_i^{(j)}}.
\end{eqnarray}

\vspace{-2.0mm}
Using the above procedure, we update {\small $P(b_i^{(j)}=+1)$} and 
{\small $P(b_i^{(j)}=-1)$} for all $i=0,\cdots,{\small 2k-1}$ and 
$j=0,\cdots,q-1$ sequentially. This completes one iteration of the 
algorithm; i.e., each iteration involves the computation of 
$\alpha_i^{(j)}$ and equations (\ref{Exp1}), 
(\ref{Exm1}), (\ref{covpm2}), (\ref{ext_fin1}), (\ref{llr_i_j_k}), 
(\ref{appx1}), and (\ref{appx2}) 
for all $i,j$. The updated values of $P(b_i^{(j)}=+1)$ 
and $P(b_i^{(j)}=-1)$ in (\ref{appx1}) and (\ref{appx2}) 
for all $i,j$ are fed back to the next iteration\footnote{The 
computation of the statistics of a current bit in an iteration
makes use of the newly computed statistics of its previous bits
(as per the ordered sequence of statistic updates)
in the same iteration and the statistics of its next bits 
available from the previous iteration.}. 
The algorithm terminates after a certain number of such iterations. 
At the end of the last iteration, hard decision is made on the final 
statistics to obtain the bit estimate $\widehat{b}_i^{(j)}$ as
$+1$ if $\Lambda_i^{(j)} \geq 1$, and $-1$ otherwise.
In coded systems, $\Lambda_{i}^{(j)}$'s are fed as 
soft inputs to the decoder.

\vspace{-2mm}
\subsection{Complexity Reduction }
\label{complex}
\vspace{-1.0mm}
The most computationally expensive operation in computing 
$\beta_{i}^{(j)}$ is the evaluation of the inverse of the covariance 
matrix, ${\bf C}_{i}^{j}$, of size $2N_rp\times 2N_rp$ 
which requires $O(N_r^3p^3)$ complexity, which can be 
reduced as follows.  Define matrix ${\bf D}$ as
\begin{eqnarray}
\label{Ckdef}
\hspace{-6mm}
{\bf D} & \hspace{-1mm} \Define & \hspace{-1mm} \sigma^2 {\bf I} + \hspace{-1mm} \sum_{l=0}^{2k-1} \sum_{m=0}^{q-1} {\bf h}_{ql+m} {\bf h}_{ql+m}^T 4 p_{l}^{m+} (1 - p_{l}^{m+}).
\end{eqnarray}
At the start of the algorithm, with $p_{i}^{j+}$ and $p_i^{(j)}$
initialized to 0.5 for all $i,j$, {\bf D} becomes
$\sigma^2 {\bf I} + {\bf H}{\bf H}^T$.

\vspace{1mm}
{\em Computation of \hspace{0.5mm}${\bf D}^{-1}$:}
We note that when the statistics of $b_{i}^{(j)}$ is updated using 
(\ref{appx1}) and (\ref{appx2}), the ${\bf D}$ matrix in (\ref{Ckdef}) 
also changes. A straightforward inversion of this updated ${\bf D}$ matrix
would require $O(N_r^3p^3)$ complexity. However, we can obtain the 
${\bf D}^{-1}$ from the previously available ${\bf D}^{-1}$ in $O(N_r^2p^2)$ 
complexity as follows. Since the statistics of
only $b_i^{(j)}$ is updated, the new ${\bf D}$ matrix is just a rank one
update of the old ${\bf D}$ matrix. Therefore, using the matrix 
inversion lemma, the new ${\bf D}^{-1}$ can be obtained from the old
${\bf D}^{-1}$ as 
\begin{eqnarray}
\label{Ckcorr}
{\bf D}^{-1} & \leftarrow & {\bf D}^{-1} - \frac{ {\bf D}^{-1}{\bf h}_{ni+j}{\bf h}_{ni+j}^T {\bf D}^{-1} } { {\bf h}_{ni+j}^T{\bf D}^{-1} {\bf h}_{ni+j} + \frac{1}{\eta} },  
\end{eqnarray}
where

\vspace{-8mm}
\begin{eqnarray}
\label{eta}
{\eta} & = & 4 p_{i}^{j+} \big(1 - p_{i}^{j+} \big) - 4 p_{i,old}^{j+} \big(1 - p_{i,old}^{j+} \big),
\end{eqnarray}
where $p_i^{j+}$ and $p_{i,old}^{j+}$ are the new $\big($i.e., after the 
update in (\ref{appx1})$\big)$ and (\ref{appx2}) $\big)$ and old (before 
the update) values, respectively. It can be seen that both the numerator and 
denominator in the 2nd term on the RHS of (\ref{Ckcorr}) can be computed in 
$O(N_r^2p^2)$ complexity. Therefore, the computation of the new ${\bf D}^{-1}$ 
using the old ${\bf D}^{-1}$ can be done in $O(N_r^2p^2)$ complexity.

{\em Computation of $({\bf C}_i^j)^{-1}$:}
Using (\ref{Ckdef}) and (\ref{covpm2}),
we can write ${\bf C}_{i}^{j}$ in terms of ${\bf D}$ as
\begin{eqnarray}
\label{cinv_cmp1}
{\bf C}_{i}^{j} & = & {\bf D} - 4 p_{i}^{j+} (1 - p_{i}^{j+}) \,{\bf h}_{qi+j} \, {\bf h}_{qi+j}^T. 
\end{eqnarray} 
We can compute $({\bf C}_{i}^{j})^{-1}$ from ${\bf D}^{-1}$ at a 
reduced complexity using the matrix inversion lemma, which states that

\vspace{-4mm}
{\small
\begin{eqnarray}
\label{cinv_cmp2}
({\bf P} + {\bf Q}{\bf R}{\bf S})^{-1}  = \, {\bf P}^{-1}  - {\bf P}^{-1} {\bf Q} ( {\bf R}^{-1} + {\bf S}{\bf P}^{-1}{\bf Q} )^{-1} {\bf S}{\bf P}^{-1}.
\end{eqnarray}
}

\vspace{-6mm}
Substituting ${\bf P}_{2N_rp\times 2N_rp} = {\bf D}$, 
${\bf Q}_{2N_rp\times 1} = {\bf h}_{qi+j}$, 
${\bf R}_{1\times 1} = - 4 p_{i}^{j+}(1 - p_{i}^{j+})$, and 
${\bf S}_{1\times 2N_rp} = {\bf h}_{qi+j}^T$ in (\ref{cinv_cmp2}), 
we get
\begin{eqnarray}
\label{cinv_cmp3}
\hspace{-2mm}
({\bf C}_{i}^{j})^{-1} \,\, = \,\, {\bf D}^{-1} - \frac{ {\bf D}^{-1} \, {\bf h}_{qi+j} \, {\bf h}_{qi+j}^T \, {\bf D}^{-1} } { {\bf h}_{qi+j}^T \, {\bf D}^{-1} \, {\bf h}_{qi+j} - \frac { 1} { 4p_{i}^{j+}(1 - p_{i}^{j+})} },  
\end{eqnarray}
which can be computed in $O(N_r^2p^2)$ complexity.

{\em Computation of ${\mbox{\boldmath{$\mu$}}}_i^{j+}$ and 
${\mbox{\boldmath{$\mu$}}}_i^{j-}$:} 
Computation of $\beta_{i}^{(j)}$ involves the computation of 
${\mbox{\boldmath{$\mu$}}}_i^{j+}$ and ${\mbox{\boldmath{$\mu$}}}_i^{j-}$
also.
From (\ref{Exm1}), it is clear that ${\mbox{\boldmath{$\mu$}}}_i^{j-}$ 
can be computed from ${\mbox{\boldmath{$\mu$}}}_i^{j+}$ with a 
computational overhead of only $O(N_rp)$. From (\ref{Exp1}), it can be
seen that computing ${\mbox{\boldmath{$\mu$}}}_i^{j+}$ would require 
$O(qN_rpk)$ complexity. However, this complexity can be reduced as 
follows. Define vector ${\bf u}$ as 
\begin{eqnarray}
\label{Ekdef}
{\bf u} & \Define & \sum_{l=0}^{2k-1} \sum_{m=0}^{q-1} {\bf h}_{ql+m} \big(2p_{l}^{m+} - 1\big). 
\end{eqnarray}
Using (\ref{Exp1}) and (\ref{Ekdef}), we can write
\begin{eqnarray}
\label{Ekdef2}
{\mbox{\boldmath{$\mu$}}}_i^{j+} & = & {\bf u}+2\big(1 - p_{i}^{j+}\big){\bf h}_{qi+j}.
\end{eqnarray}
{\bf u} can be computed iteratively at $O(N_rp)$ complexity as follows. 
When the statistics of $b_{i}^{(j)}$ is updated, we can obtain the new
${\bf u}$ from the old ${\bf u}$ as
\begin{eqnarray}
\label{Ekcorr}
{\bf u} & \leftarrow & {\bf u} +  2\big( p_{i}^{j+} - p_{i,old}^{j+}\big) {\bf h}_{ni+j},
\end{eqnarray}
whose complexity is $O(N_rp)$. Hence, the computation of 
${\mbox{\boldmath{$\mu$}}}_i^{j+}$ in (\ref{Ekdef}) 
and ${\mbox{\boldmath{$\mu$}}}_i^{j-}$ in (\ref{Ekdef2}) 
needs $O(N_rp)$ complexity. The listing of the proposed PDA algorithm is
summarized in the Table-I in the next page.

\newpage
\hrule
{\small
\vspace{-0.25mm}
Table-I: Proposed PDA-based Algorithm Listing
\vspace{0.5mm}
\hrule
\vspace{1mm}
\hspace{3mm} {\em Initialization} 

\vspace{-1mm}
1. $p_{i}^{j+} = p_{i}^{j-} = 0.5$, \hspace{1mm} 
$\Lambda_{i}^{(j)} = 1$, 

\vspace{-1mm}
\hspace{20mm} $\forall i=0,1,\cdots 2k-1, j=0,1,\cdots,q-1$. 

\vspace{-1mm}
2. ${\bf u} = {\bf 0}, \hspace{1mm} 
{\bf D}^{-1}=\big({\bf H}{\bf H}^T + \sigma^2 {\bf I}\big)^{-1}$. 

\vspace{-1mm}
3. $num\_iter$: number of iterations

\vspace{-1mm}
4. $\kappa = 1$; \hspace{6mm} $\kappa$ is the iteration number 

\vspace{-1mm}
\hspace{3mm} {\em Statistics update in the $\kappa$th iteration}

\vspace{-1mm}
5. for $i$ = 0 to $2k-1$ 

\vspace{-1mm}
6. for $j$ = 0 to $q-1$ 

\vspace{-1mm}
\hspace{3mm} {\em Update of statistics of bit $b_i^{(j)}$} 

\vspace{-1mm}
7. ${\mbox{\boldmath{$\mu$}}}_i^{j+} \, = \, {\bf u}+2\big(1-p_{i}^{j+}\big){\bf h}_{qi+j}$

\vspace{-1mm}
8. ${\mbox{\boldmath{$\mu$}}}_i^{j-} \, = \, {\mbox{\boldmath{$\mu$}}}_i^{j+} - 2{\bf h}_{qi+j}$

9. $({\bf C}_{i}^{j})^{-1} \, = \, {\bf D}^{-1} - \frac{ {\bf D}^{-1} \, {\bf h}_{qi+j} \, {\bf h}_{qi+j}^T \, {\bf D}^{-1} } { {\bf h}_{qi+j}^T \, {\bf D}^{-1} \, {\bf h}_{qi+j} - \frac { 1} { 4p_{i}^{j+}(1 - p_{i}^{j+})} }$

\vspace{-1mm}
10. ${\small \beta_{i}^{j} \, = \, e^{-\left( ({\bf y} - {\mbox{\boldmath{$\mu$}}}_{i}^{j+})^T ({\bf C}_{i}^{{j}})^{-1} ({\bf y} - {\mbox{\boldmath{$\mu$}}}_{i}^{j+}) - ({\bf y} - {\mbox{\boldmath{$\mu$}}}_{i}^{j-})^T ({\bf C}_{i}^{j})^{-1} ({\bf y} - {\mbox{\boldmath{$\mu$}}}_{i}^{j-})\right )} }$

\vspace{-1mm}
11. $p_{i,old}^{j+}\,=\,p_{i}^{j+}$, \hspace{4mm} $p_{i,old}^{j-}\,=\,p_{i}^{j-}$ 

\vspace{-1mm}
12. $\alpha_i^{(j)} = \frac{p_{i,old}^{j+}}{p_{i,old}^{j-}}$ 

\vspace{-1mm}
13. $\Lambda_i^{(j)} = \beta_i^{(j)} \alpha_i^{(j)}$ 

\vspace{-1mm}
14. $p_i^{j+} \, = \, \frac{\Lambda_i^{(j)}}{1+\Lambda_i^{(j)}}$,
    \hspace{4mm} $p_i^{j-} \, = \, \frac{1}{1+\Lambda_i^{(j)}}$

\vspace{-1mm}
\hspace{3mm} {\em Update of ${\bf u}$ and ${\bf D}^{-1}$} 

\vspace{-1mm}
15. ${\bf u} \, \leftarrow \, {\bf u} +  2\big( p_{i}^{j+} - p_{i,old}^{j+}\big) {\bf h}_{qi+j}$

\vspace{-1mm}
16. ${\eta} \, = \, 4 p_{i}^{j+} \big(1 - p_{i}^{j+} \big) - 4 p_{i,old}^{j+} \big(1 - p_{i,old}^{j+} \big)$

\vspace{-1mm}
17. ${\bf D}^{-1} \, \leftarrow \, {\bf D}^{-1} - \frac{ {\bf D}^{-1}{\bf h}_{qi+j}{\bf h}_{qi+j}^T {\bf D}^{-1} } { {\bf h}_{qi+j}^T{\bf D}^{-1} {\bf h}_{qi+j} + \frac{1}{\eta} }$

\vspace{-1mm}
18. end; \hspace{6mm} End of for loop starting at line 5 

\vspace{-1mm}
19. if ($\kappa = num\_iter$) goto line 21 

\vspace{-1mm}
20. $\kappa = \kappa+1$, goto line 5 

\vspace{-1mm}
21. ${\widehat b}_i^{(j)} = \mbox{sgn}\big(\log(\Lambda_{i}^{(j)})\big)$ 

\vspace{-1mm}
\hspace{15mm} $\forall i=0,1,\cdots,2k-1, \hspace{1mm} j=0,1,\cdots,q-1$

\vspace{-1mm}
22. ${\widehat x}_i = \sum_{j=0}^{q-1} 2^j \, {\hat b}_i^{(j)}, \hspace{3mm} \forall i=0,1,\cdots,2k-1$ 

\vspace{-1mm}
23. Terminate 
\vspace{2mm}
}
\hrule

\vspace{-2.0mm}
\subsection{Overall Complexity }
\label{overall}
\vspace{-1mm}
We need to compute ${\bf H}{\bf H}^T$ at the start of the algorithm. This 
requires $O(qkN_r^2p^2)$ complexity. So the computation of the initial 
${\bf D}^{-1}$ in line 2 requires $O(qkN_r^2p^2)+O(N_r^3p^3)$. 
Based on the complexity reduction in Sec. \ref{complex}, 
the complexity in updating the statistics of one constituent bit 
(lines 7 to 17) is $O(N_r^2p^2)$. So, the complexity for the update of 
all the $2qk$ constituent bits in an iteration is $O(qkN_r^2p^2)$. 
Since the number of iterations is fixed, the overall complexity of the 
algorithm is $O(qkN_r^2p^2)+O(N_r^3p^3)$. For $N_t=N_r$, since there 
are $k$ symbols per STBC and $q$ bits per symbol, the overall 
complexity per bit is $O(p^2N_t^2)$. 

\vspace{-2.0mm}
\section{Results and Discussions}
\label{sec4}
\vspace{-1.0mm}
In this section, we present the simulated uncoded/coded BER of 
the PDA algorithm in decoding non-orthogonal STBCs from CDA\footnote{Our
simulation results showed that the performance of FD-ILL 
($\delta=e^{\sqrt{5}{\bf j}}, t=e^{-{\bf j}}$) and ILL ($\delta=t=1$) 
STBCs with PDA
decoding were almost the same. Here, we present the performance of
ILL STBCs.}. Number of iterations in the PDA
algorithm is set to $m=10$ in all the simulations.

\begin{figure}
\epsfxsize=8.5cm
\hspace{-3mm}
\epsfbox{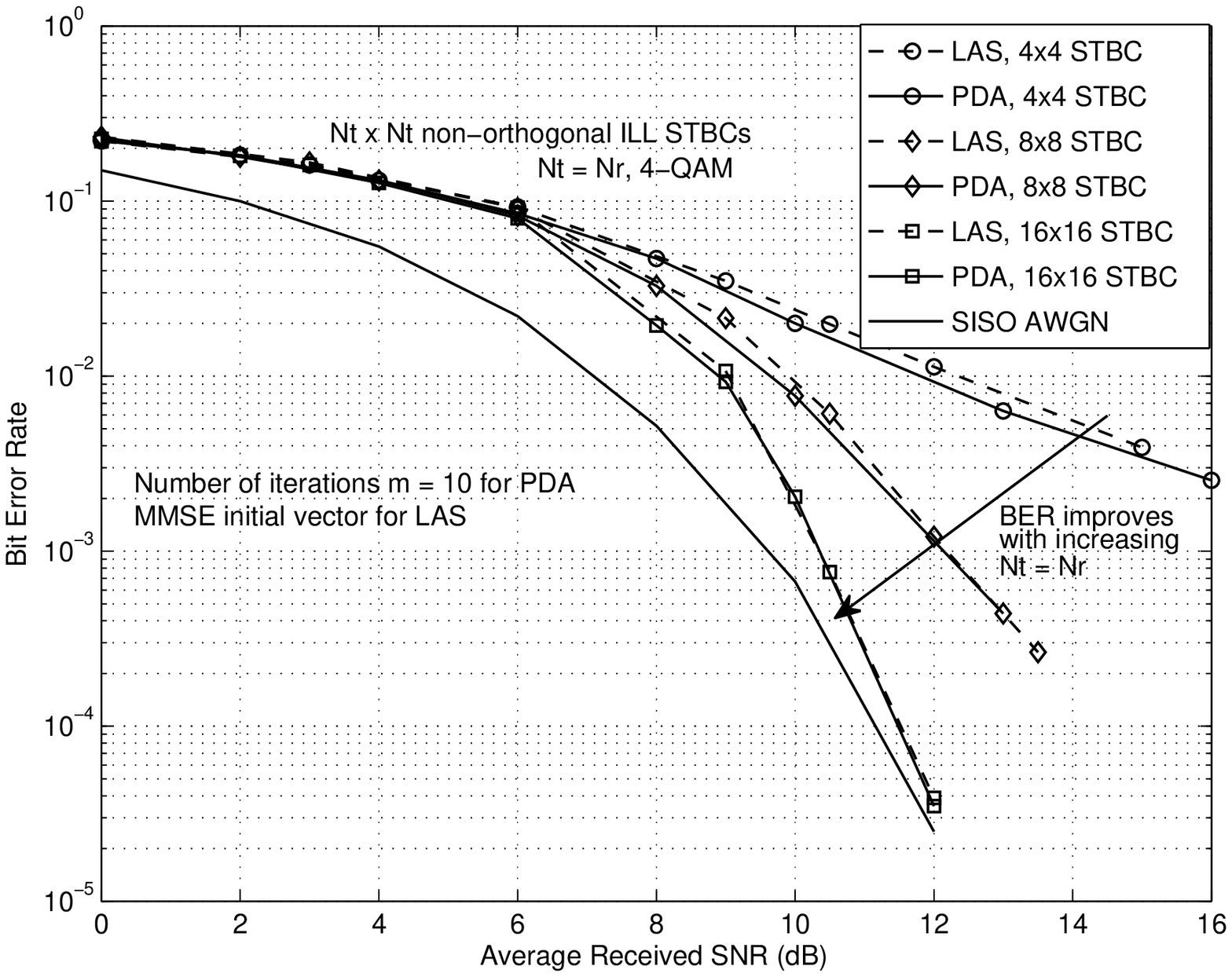}
\vspace{-1.5mm}
\caption{Comparison of uncoded BER of PDA and LAS algorithms 
in decoding $4\times 4$, $8\times 8$, $16\times 16$ ILL STBCs. 
$N_t=N_r$, 4-QAM. 
\# iterations $m=10$ for PDA. MMSE initial vector for LAS. 
{\em BER improves for increasing STBC sizes.
With 4-QAM, PDA and LAS algorithms achieve almost the same performance.}
}
\vspace{-3.0mm}
\label{fig1}
\end{figure}

{\em PDA versus LAS performance with 4-QAM:}
In Fig. \ref{fig1}, we plot the uncoded BER of the PDA algorithm as a 
function of average received SNR per rx antenna, $\gamma$, in decoding 
{\small $4\times 4$, $8\times 8$, $16\times 16$} STBCs from CDA with 
{\small $N_t=N_r$} and {\small 4-QAM}. Perfect channel state information 
at the receiver (CSIR) and i.i.d fading are assumed. For the same settings, 
the performance of the LAS algorithm in \cite{jsac}-\cite{gcom08} with 
MMSE initial vector are 
also plotted for comparison. From Fig. \ref{fig1}, it is seen that 
\begin{itemize}
\item the BER performance of PDA algorithm improves and approaches SISO 
AWGN performance as $N_t=N_r$ is increased; e.g., performance close to 
within about 1 dB from SISO AWGN performance is achieved at $10^{-3}$ 
uncoded BER in decoding $16\times 16$ STBC from CDA having 512 real 
dimensions, and this illustrates the ability of the PDA algorithm to 
achieve excellent performance at low complexities in large non-orthogonal 
STBC MIMO. 
\item
with 4-QAM, PDA and LAS algorithms achieve almost the same performance.
\end{itemize}

\begin{figure}
\vspace{-4.0mm}
\epsfxsize=8.5cm
\hspace{-3mm}
\epsfbox{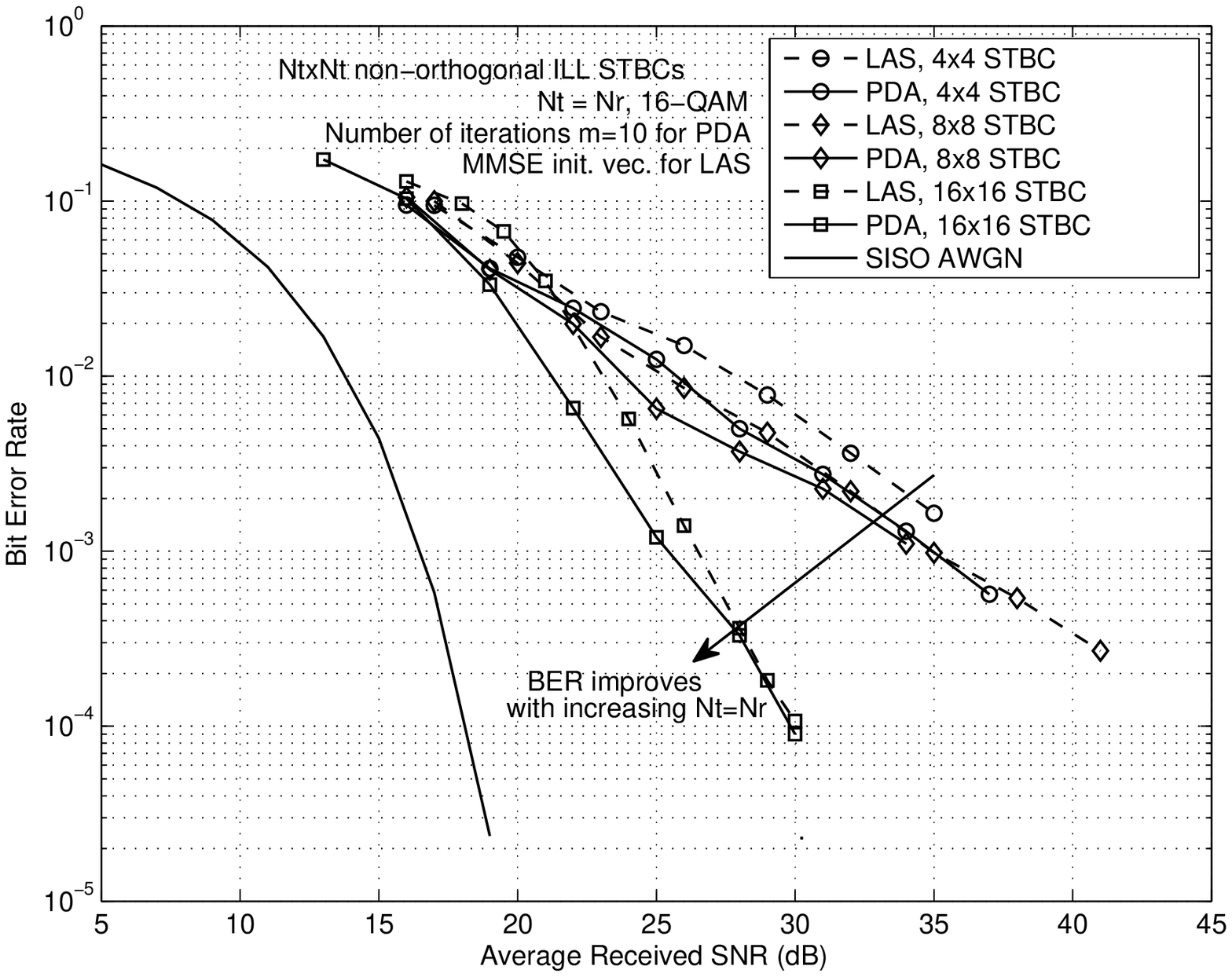}
\vspace{-1.0mm}
\caption{Comparison of uncoded BER of PDA and LAS algorithms
in decoding $4\times 4$, $8\times 8$, $16\times 16$ ILL STBCs. 
$N_t=N_r$, 16-QAM. 
\# iterations $m=10$ for PDA. MMSE initial vector for LAS. 
{\em With 16-QAM, PDA performs better than LAS at low SNRs.} 
}
\vspace{-4.0mm}
\label{fig2}
\end{figure}

{\em PDA versus LAS performance with 16-QAM:}
Figure \ref{fig2} presents an uncoded BER comparison between PDA and
LAS algorithms for $16\times 16$ STBC from CDA with $N_t=Nr=16$
and 16-QAM under perfect CSIR and i.i.d fading. It can be seen that 
the PDA algorithm performs better at low SNRs than the LAS algorithm. 
For example, with $8\times 8$ and $16\times 16$ STBCs, at low SNRs 
(e.g., $< 25$ dB for $16\times 16$ STBC), PDA algorithm performs 
better by about 1 dB compared to LAS algorithm at $10^{-2}$ uncoded BER.

\begin{figure}
\hspace{-3mm}
\epsfxsize=8.50cm
\epsfbox{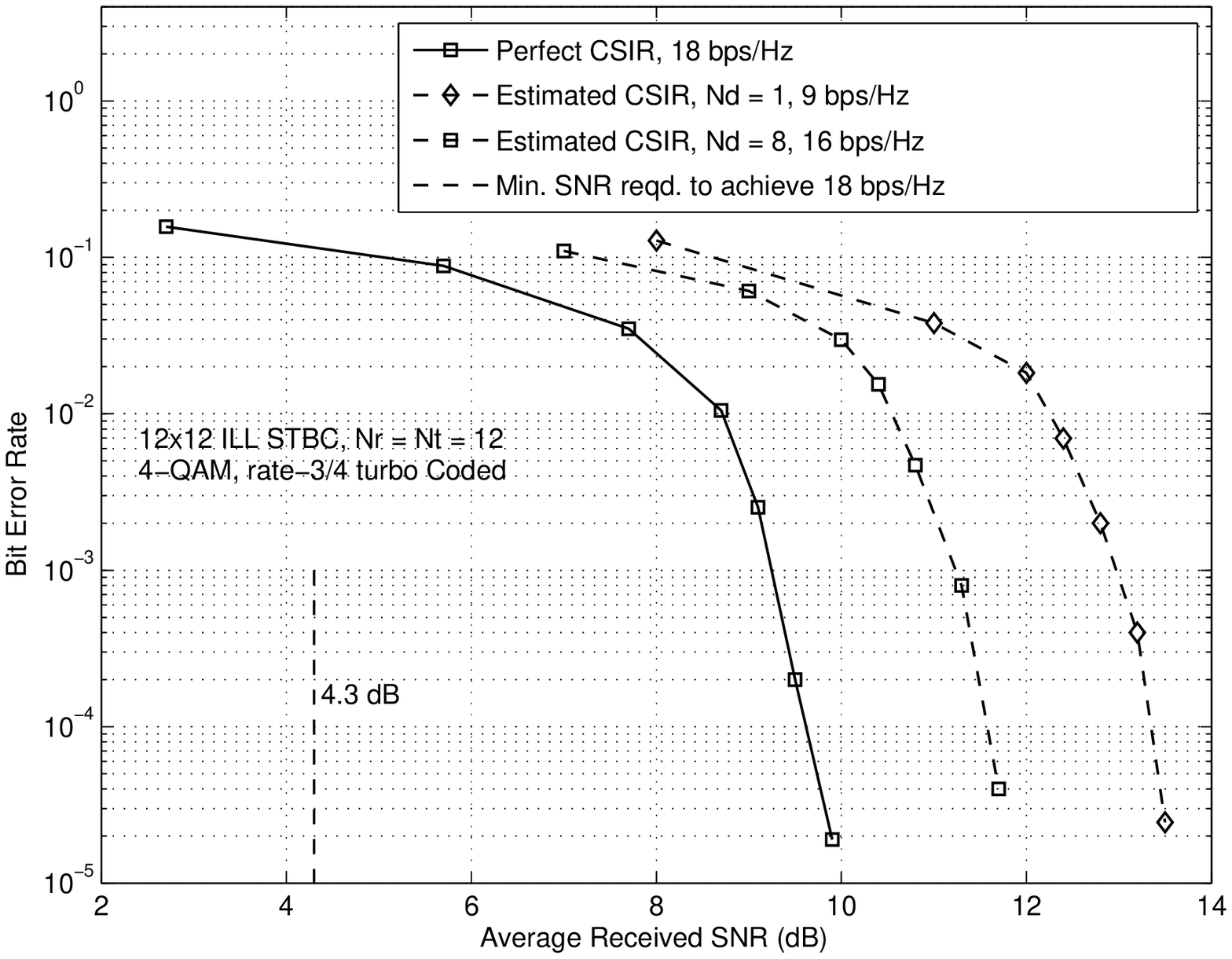}
\vspace{-1.0mm}
\caption{Turbo coded BER of the PDA algorithm in decoding $12\times 12$ 
ILL STBC with $N_t=N_r$, 4-QAM, rate-3/4 turbo code, 
18 bps/Hz and $m=10$ for $i)$ perfect CSIR, and $ii)$ estimated
CSIR using 2 iterations between PDA decoding/channel estimation. 
{\em With perfect CSIR, PDA performs close to within about 5 dB from 
capacity. With estimated CSIR, performance approaches to that with 
perfect CSIR with increasing coherence times.}
}
\vspace{-1mm}
\label{fig3}
\end{figure}

{\em Turbo coded BER performance of PDA:}
Figure \ref{fig3} shows the rate-3/4 turbo coded BER of the PDA algorithm 
under perfect CSIR and i.i.d fading for $12\times 12$ ILL STBC with 
$N_t\hspace{-0.5mm}=\hspace{-0.5mm}N_r\hspace{-0.5mm}=\hspace{-0.5mm}12$ 
and 4-QAM, which corresponds to a spectral efficiency of 18 bps/Hz. The
theoretical minimum SNR required to achieve 18 bps/Hz spectral efficiency 
on a $N_t\hspace{-0.5mm}=\hspace{-0.5mm}N_r\hspace{-0.5mm}=\hspace{-0.5mm}12$ 
MIMO channel with perfect CSIR and i.i.d fading is 4.3 dB (obtained 
through simulation of the ergodic capacity formula \cite{jafarkhani}). 
From Fig. \ref{fig3}, it is seen that the PDA algorithm is able to 
achieve vertical fall in coded BER within about 5 dB from the 
theoretical minimum SNR, which is a good nearness to capacity performance. 

\begin{figure}
\vspace{-0mm}
\centering
\includegraphics[width=3.2in]{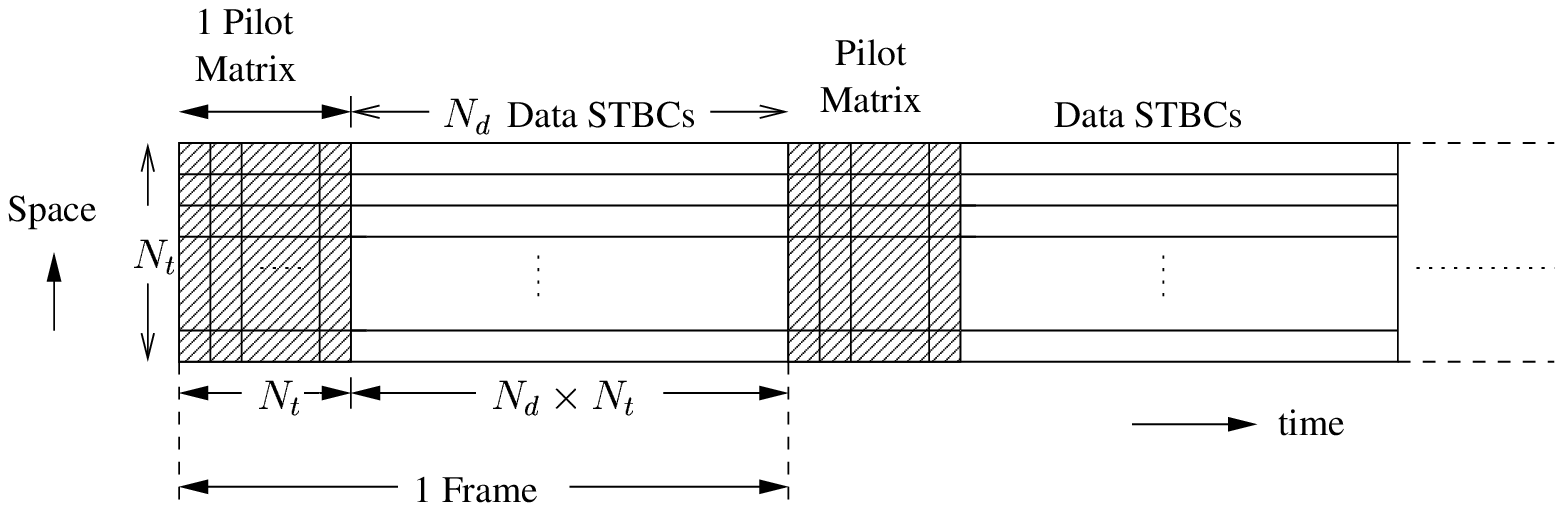}
\vspace{-0mm}
\caption{Transmission scheme with one pilot matrix followed by
$N_d$ data STBC matrices in each frame. }
\label{fig4}
\vspace{-0.0mm}
\end{figure}

\vspace{1mm}
{\em Iterative PDA Decoding/Channel Estimation:}
We relax the perfect CSIR assumption by considering a training based 
iterative PDA decoding/channel estimation scheme. Transmission is 
carried out in frames, where one $N_t\times N_t$ pilot matrix 
(for training purposes) followed by $N_d$ data STBC matrices are sent 
in each frame as shown in Fig. \ref{fig4}. One frame length, $T$, 
(taken to be the channel coherence time) is $T=(N_d+1)N_t$ channel 
uses. The proposed scheme works as follows \cite{zaki}: $i)$ obtain an 
MMSE estimate of the channel matrix during the pilot phase, $ii)$ use the 
estimated channel matrix to decode the data STBC matrices using PDA 
algorithm, and $iii)$ iterate between channel estimation and PDA 
decoding for a certain number of times. For $12\times 12$ STBC from 
CDA, in addition to perfect CSIR performance, Fig. \ref{fig3} 
also shows the performance with CSIR estimated using the proposed 
iterative decoding/channel estimation scheme for $N_d=1$ and $N_d=8$. 
2 iterations between decoding and channel estimation are used. With 
$N_d=8$ (which corresponds to large coherence times, i.e., slow fading) 
the BER and bps/Hz with estimated CSIR get closer to those with 
perfect CSIR.

\begin{figure}
\hspace{-1mm}
\epsfxsize=8.5cm
\epsfbox{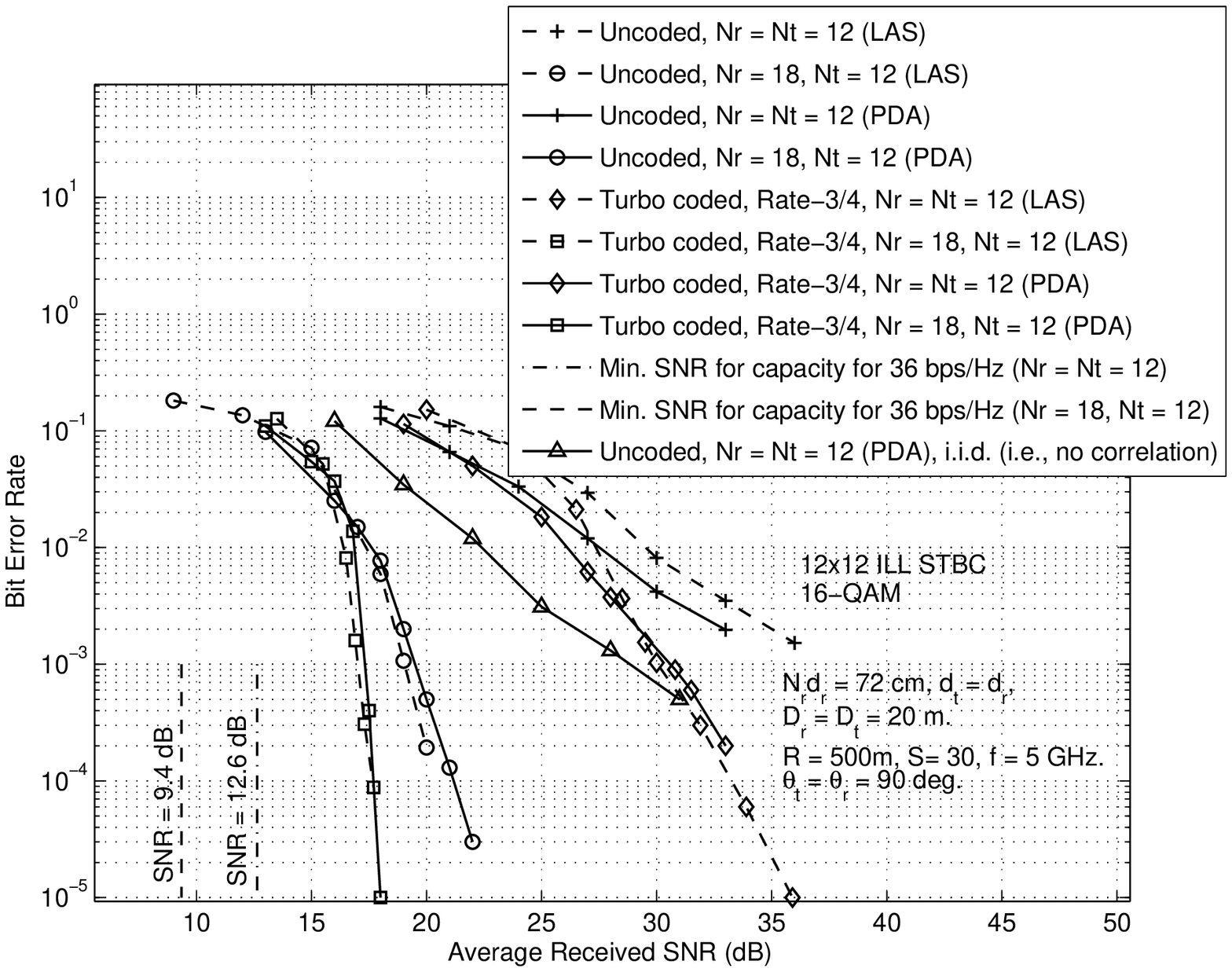}
\vspace{-1.0mm}
\caption{Effect of spatial correlation on the performance of PDA in
decoding $12 \times 12$ STBC from CDA. $N_t=12$, $N_r=12,18$, 16-QAM,
rate-3/4 turbo code, 36 bps/Hz. Correlated channel 
parameters: $f_c=5$ GHz, $R=500$ m, $S=30$, $D_t=D_r=20$ m, 
$\theta_t=\theta_r=90^\circ$, $N_rd_r=72$ cm, $d_t=d_r$. 
{\em Correlation degrades performance; using $N_r > N_t$ alleviates 
the this performance loss. }
}
\vspace{-3.0mm}
\label{fig5}
\end{figure}

{\em Effect of Spatial MIMO Correlation:}
In Figs. \ref{fig1} to \ref{fig3}, we assumed i.i.d fading. But spatial 
correlation at transmit/receive antennas and the structure of scattering 
and propagation environment can affect the rank structure of the MIMO 
channel resulting in degraded performance \cite{mimo1}.
We relaxed the i.i.d. fading assumption by considering the correlated
MIMO channel model in \cite{gesbert}, which takes into 
account carrier frequency ($f_c$), spacing between antenna elements 
{\small ($d_t,d_r$)}, distance between tx and rx antennas ($R$), and 
scattering environment. In Fig. \ref{fig5}, we plot the BER of the PDA 
algorithm in decoding {\small $12\times 12$} STBC from CDA with perfect 
CSIR in $i)$ i.i.d. fading, and $ii)$ correlated MIMO fading model in 
\cite{gesbert}. It is seen that, compared to i.i.d fading, there is a 
loss in diversity order in spatial correlation for $N_t=N_r=12$; 
further, use of more rx antennas ($N_r=18, N_t=12$) alleviates this loss 
in performance. We can decode {\em perfect codes} \cite{perf06},\cite{perf07}
of large dimensions also using the proposed PDA algorithm. 

\end{document}